\documentclass[12pt,preprint]{aastex}

\shorttitle{The  acoustic cutoff frequency of the Sun and the activity cycle }

\shortauthors{Jim\'enez, Garc\'\i a \& Pall\'e}

\begin{document}
\title{The  acoustic cut-off frequency of the Sun and the solar magnetic activity cycle }
\author{A. Jim\'enez\altaffilmark{1,2}, R A Garc\'\i a\altaffilmark{3}, P L Pall\'e\altaffilmark{1,2} }
\altaffiltext{1}{Instituto de Astrof\'isica de Canarias, E-38205 La Laguna, Tenerife, Spain}
\altaffiltext{2} {Departamento de Astrof\'isica, Universidad de La Laguna, E-38206 La Laguna, Tenerife, Spain}
\altaffiltext{3} {Laboratoire AIM, CEA/DSM-CNRS, Universit\'e Paris 7 Diderot, IRFU/Sap, Centre de Saclay, F-91191 Gif-sur-Yvette, France}

\begin{abstract}
The acoustic cut-off frequency---the highest frequency for acoustic solar eigenmodes---is an important parameter of 
the solar atmosphere as it determines the upper boundary of the p-mode resonant cavities. At frequencies beyond
this value, acoustic disturbances are no longer trapped  but traveling waves. 
Interference amongst them give rise to higher-frequency peaks---the pseudomodes---in the 
solar acoustic spectrum. The pseudomodes are shifted slightly in frequency with respect to 
p-modes making possible the use of pseudomodes to determine the acoustic cut-off frequency.  Using data from 
GOLF and VIRGO instruments on board the SOHO spacecraft, we calculate the acoustic cut-off 
frequency using the coherence function between both the velocity and intensity sets of data. 
By using data gathered by these instruments during the entire lifetime of the mission (1996 
till the present), a variation in the acoustic cut-off frequency with the solar magnetic activity cycle is found.
\end{abstract}

\maketitle

\section{Introduction}
Solar p-modes are essentially resonant acoustic waves that can be regarded as a superposition of 
outward and inwardly propagating waves that interfere constructively. At certain discrete 
frequencies the interference is maximally constructive yielding  the eigenfrequencies of the 
acoustic cavity within which the p-modes propagate. The lower boundary of these resonant cavities lies 
at a depth inside the Sun at which the horizontal phase speed of the wave equals the local sound 
speed. The upper boundary of the p-mode cavities lies near the solar surface and its location 
depends on the frequency and wavenumber of the modes. For frequencies higher than a certain value, called the acoustic cut-off frequency ($\nu_{\rm ac}$), acoustic disturbances are no longer trapped and propagate as traveling waves through the 
chromosphere to the base of the corona. These high-frequency peaks (hereafter, pseudomodes) show a 
clear  periodic structure beyond  $\nu_{\rm ac}$ (Jefferies et al.\ 1988; Libbrecht 1988; Duvall et al.\  1991;
Garc\'\i a et al.\  1998; Chaplin et al. \ 2002; Jim\'enez et al.  2005; Jim\'enez 2006). Several models have been
proposed to explain the nature of the pseudomodes (Balmforth \& Gough 1990; Kumar et al.\ 1990; Kumar 1994; Jain \& Roberts 
1996; Garc\'\i a et al.\ 1998). Nowadays, it is generally believed that 
the full-disk integrated (low-degree) pseudomodes  arise from  geometric 
interference between direct waves emitted from a subphotospheric source and indirect waves produced by partial 
reflection on the far side of the Sun (Garc\'\i a et al.\ 1998). For higher-degree modes the indirect waves are those
emitted downwards toward the solar interior and refracted back to the solar surface (Kumar et al.\ 1990; Kumar 1994). 

The acoustic cut-off frequency is given by  $\nu_{\rm ac}$= $c/2H_{\rho}$ $\propto$ $g/\sqrt{T_{eff}}$ $\propto{MR^{-2}T_{eff}^{-1/2}}$, $H_{\rho}$ being the density scale height, 
c the sound speed, g the gravitational field, M the mass, R the radius, and $T_{eff}$ the temperature at the photosphere.  The acoustic cut-off frequency
sets the borderline between p-modes and pseudomodes and it is related with other parameters of the oscillation spectrum, for example, 
$\nu_{\rm max}$, the frequency of maximum power of the oscillations. It was conjectured by Brown et al. (1991) that the 
$\nu_{\rm max}$ scales the cut-off frequency and the increasing number of observations of solar-like stars has confirmed this relation
(Beeding and Kjeldsen, 2003; Stello et al. 2009). $\nu_{\rm max}$ is associated with the coupling between turbulent convection and oscillations and results from a balance between the damping and the driving of the modes. Asteroseismic observations suggest that  $\nu_{\rm max}$ corresponds to the plateau of the damping rates (e.g., Benomar et al. 2009; Barban et al. 2009; Deheuvels et al. 2010; Ballot et al. 2011). Moreover, Belkacem et al. (2011) has addressed the question of whether the depression of the damping rates determines $\nu_{\rm max}$  because there is a resonance between thermal time-scale in the superadiabatic region and the modal
period, implying that 
$\nu_{\rm max}$ does not scale only with $\nu_{\rm ac}$ but also with the Mach number. In other stars for which stellar parameters are well known, the relation between $\nu_{\rm max}$  and $\nu_{\rm ac}$ could directly give the value of the Mach number in the uppermost convective layers.

 The intrinsic importance of $\nu_{\rm ac}$ increases with asteroseismology because the simple scaling relations for asteroseismic quantities have proven very useful when analyzing stellar oscillations, and these scaling relations use the Sun as a reference (Kjeldsen \& Beeding 1995). For example, it is well established that to a good approximation $\Delta\nu_{n,\ell}$, the so-called large frequency separation between consecutive overtones, is proportional to the square root of the stellar density (e.g., Ulrich 1986):

  \begin{equation}
 \frac{\Delta\nu_{n,\ell}}{\Delta\nu_{n,\ell,\sun}}=\sqrt{\frac{\rho}{\rho_{\sun}}}=\frac{(M/M_{\sun})^{0.5}(T_{eff}/T_{eff,\sun})^3}{(L/L_{\sun})}
 \end{equation}

\noindent and also, following Brown et al. (1991) and Kjeldsen \& Bedding (1995), we expect $\nu_{\rm max}$ to scale as $\nu_{\rm ac}$:

 \begin{equation}
 \frac{\nu_{max}}{\nu_{max,\sun}}= \frac{\nu_{ac}}{\nu_{ac,\sun}}= \frac{M/M_{\sun}(T_{eff}/T_{eff,{\sun}})^{3.5}}{L/L_{\sun}}
 \end{equation}

For the Sun  $\nu_{ac,\sun}$ $\simeq 1.7$ $\nu_{\rm max,\sun}$ (Balmforth and Gough 1990; Fossat et al. 1992). The theoretical value inferred is $\nu_{\rm ac,\sun}$ $\simeq 5300$  $\mu$Hz.

Observationally some estimates have been obtained in the past (Claverie et al. 1981; Pall\'e et al. 1986, 1992; Duval et al. 1991; Fossat et al. 1992) yielding different results in the range  5300--5700 $\mu$Hz. Before the discovery of pseudomodes it might have been expected that a way to measure $\nu_{\rm ac}$ would be to look for a sudden drop in the power density signal on the high-frequency side of the spectrum. All the previous observational results were obtained before the discovery of pseudomodes and probably the weak signal of the first pseudomodes were considered as p-modes. Although, in principle, the existence of pseudomodes does complicate the determination of the acoustic cut-off frequency, this in fact helps to determine it. In fact, when pseudomodes are taken into account, a lower value is obtained: $\nu_{\rm ac}$ $\simeq$ 5100 $\mu$Hz (Jim\'enez 2006).

The method of measuring $\nu_{ac}$ in this research is as follows. In the acoustic spectrum of the Sun, 
the large frequency separation between consecutive modes of the same degree 
$\Delta\nu_{n,\ell}$=$\nu_{\rm n,\ell}$-$\nu_{n-1,\ell}$ is approximately equal 
to the inverse of the sound travel time from the upper reflection point to the lower 
turning point and back. This $\Delta\nu_{n,\ell}$ decreases if the lower turning point 
moves inward (increasing $\nu$ or decreasing $\ell$), or if the outer reflection point 
moves outward. At a given spherical harmonic, the observed frequency spacing between peaks 
decreases with increasing frequency. However several authors (Kumar et al. 1994; Nigam 
\& Kosovichev 1996) have pointed out  that, between 5000 $\mu$Hz and 5500 $\mu$Hz, the 
frequency spacing increases slightly, this feature probably being associated with the 
acoustic cut-off frequency indicating the transition from trapped to traveling waves.

If $\Delta\nu_{n,\ell}$ increases around $\nu_{\rm ac}$, all the peaks with frequencies $\nu>\nu_{\rm ac}$ will be shifted relative to the peaks with frequencies $\nu<\nu_{\rm ac}$. 
Finding the frequency at which these  shifts takes place would provide a good measurement 
of the acoustic cut-off frequency. In the present study, the coherence function between intensity 
and velocity signals will be used instead of the power spectra to avoid  intensity contamination, 
as  will be explained in Section 4.

The transition between p-modes and pseudomodes also could be observed in other oscillation parameters. 
In particular, if the phase relations (phase shifts)  change for frequencies beyond $\nu_{\rm ac}$. 
These phase shifts can correspond to velocity--velocity or 
intensity--intensity measurements at two different spectral lines (two different formation heights), 
or to intensity--velocity measurements of spectral lines, narrow-band photometry, etc. For velocity--velocity 
(or intensity--intensity) observations, a phase difference of around 0$\degr$ should be 
expected in the frequency range corresponding to standing (trapped) waves, i.e., the p-mode range
(e.g., Fossat et al.\ 1992; Pall\'e et al.\ 1992; Jim\'enez et al.\ 1999). For waves with frequencies 
beyond $\nu_{\rm ac}$ (the traveling 
wave range) a non-zero phase difference should be expected  as was already measured by Staiger (1987), who found  
close to zero V--V phase differences between 2500 $\mu$Hz and 5000 $\mu$Hz,  but for 
frequencies between 5000 $\mu$Hz and 7000 $\mu$Hz he also found that the phase difference 
changes almost linearly, and that these values were in good agreement with theoretical values 
calculated by Schmieder (1977, 1978) under the assumption that vertically traveling waves exist 
beyond the acoustic cut-off frequency.

For intensity--velocity observations the phase differences give information about the adiabatic 
or non-adiabatic behavior of the solar atmosphere. In the adiabatic case a value of $-90\degr$ 
or $+90\degr$ (depending on the sign convention for upward/downward positive velocity)  is 
expected for the p-mode range and a value of 0$\degr$ for a model close to isothermal (Marmolino 
\& Severino 1991). For non-adiabatic conditions the phase differences change with frequency 
depending on the model used (Gough 1985; Houdek et al. 1995). The results of Jim\'enez (2002) 
show that in the p-mode range the I-V phase differences do not show an exactly adiabatic behavior
but one that is close to it. Our concern in the present investigation is that these I-V phase differences 
in the p-mode range are found to be close to $-90\degr$, are roughly constant with frequency, and 
have no dependence on solar activity  (Jim\'enez et al. 2002).

The theoretical results of Schmieder (1978) show that the I-V phase differences for frequencies 
beyond  $\nu_{\rm ac}$ also change almost linearly in the frequency range between 5000 $\mu$Hz and 
6500 $\mu$Hz because of the traveling nature of the waves. Taking into account the different I-V phase 
relationships between p-modes and pseudomodes  it would be interesting to know how I-V phase differences
are affected by the transition between standing and traveling waves and determine whether this effect
corresponds to $\nu_{\rm ac}$ .

\rm

\section{Instrumentation}
\subsection{VIRGO/SPM (Intensities)}
The Solar Photometers (SPM) of the Variability of solar IRradiance and Gravity 
Oscillations (VIRGO) package (Fr\"{o}hlich et al.\ 1995, 1997) aboard the Solar
and Heliospheric Observatory (SOHO) mission, consists of three independent 
photometers, centered around 402, 500 and 862 nm (the blue, green, and red channel 
respectively). They measure the spatially integrated solar intensity over a 5 nm bandpass 
at a one minute cadence. The data obtained by the VIRGO/SPM instrument over the 
mission have been of uniformly high quality, whether before or after the loss 
of contact with the spacecraft for several months in 1998.

\subsection{GOLF (Velocity)}

The Global Oscillations at Low Frequency (GOLF)  instrument  on board SOHO (Gabriel et al.\ 1995, 1997)
is a resonant scattering spectrophotometer that measures the line-of-sight velocity
between the Sun and the spacecraft using the sodium doublet. It uses the same 
technique as other ground-based helioseismic networks such as the International 
Research on the Interior of the Sun (IRIS, Fossat 1990) and the Birmingham Solar 
Oscillations Network (BiSON, Broomhall et al. 2009). The
GOLF window was  opened in 1996 January and became fully
operational by the end of that month. Over the following months,
occasional malfunctions in its rotating polarizing elements were
noticed that led to the decision to stop them in a predetermined
position; truly non-stop observations began by 1996 mid-April (Garc\'ia et al. 2005).
Since then, GOLF has been continuously and satisfactorily operating
in a  mode unforeseen before launch, showing fewer
limitations than anticipated. The signal, then, consists of two close
monochromatic photometric measurements in a very narrow band (25
m\AA) on a single wing of the sodium doublet (Garc\'\i a et al.\ 2005).
This signal has been calibrated as velocity and is indeed  similar in nature
to other known velocity measurements, such as those of IRIS and BiSON 
(Pall\' e et al.\ 1999). The sampling of the GOLF data used in this paper is 60 s.
Before the 1998 loss of contact with the SOHO spacecraft (June 1998), the GOLF instrument was 
operated using the blue wing of the sodium line. When thereafter contact with the 
spacecraft was reestablished (October 1998) the instrument operating mode was 
switched to the red wing of the line until November 2002. Since then, it was switched
back to the blue wing. In Table~1 we summarize the GOLF working configurations as 
well as the associated duty cycle of each series.

\begin{table}[!htb]
\caption{\label{tab:events}Time series number, dates, and sodium wing observed by GOLF in 
each period and mean duty cycle (DC) of the time series used.}
\begin{center}
\begin{tabular}{llll}
\hline
Time series & Dates Interval & Na wing&DC\\
\hline
0-19   & April 96 to June 98&blue& 91.7\% \\
20-33 & June 98  to November 2002& red&99.4\% \\
34-85 & November 2002 to May 2010 & blue&97.6\%\\
\hline
\end{tabular}
\end{center}
\end{table}

\section {Data sets}
In this study we use VIRGO/SPM (three color) intensity and GOLF velocity time 
series recorded for the last $\sim$~15 years.
A total of 85 time series of 800 consecutive days --covering the period April 
1996 till May 2010-- have been analyzed. Each one is shifted 50 days with
respect to the previous one. Each 800-day time series is later sub-divided 
into  4-day time series (5760 one-minute sampled points, with a frequency 
resolution of 2.89 $\mu$Hz). Their corresponding power spectra and the bivariate 
parameters (coherence and phase differences) are computed and the results averaged 
(time series with more than 500 missing points are excluded). Figure 1 shows a 
sample of the power spectra of the three channels of VIRGO/SPM (red, green, and 
blue), and the GOLF spectrum from 3000 $\mu$Hz (including the high frequency 
part of the 5-minute oscillations band) to the Nyquist frequency. The periodic 
structure of the pseudomodes is clearly seen in all the signals.

 To study the transition frequency range between p-modes and pseudomodes where $\nu_{\rm ac}$ 
should be located, the coherence and phase shift between the intensity and velocity signals will 
be used. In phase analyses the correct timing of the temporal series is a critical issue that 
should be carefully addressed. In the present case, VIRGO data timing is very well determined 
and remained stable throughout the mission. In contrast, GOLF timings presented some shifts during
the mission and required proper correction. The corrections introduced were satisfactory 
(see Garc\'ia et al. 2005 for more details) and time series ready to be used in the combined 
VIRGO/GOLF analysis were then produced (Jim\'enez et al. 1999; Jim\'enez 2002). In fact, additional 
tests have been performed in the present study and the resulting maximum errors in the computed I-V phase 
difference GOLF/VIRGO for the largest possible timing error (+3.3144 s) turns out to be smaller than 
8 \arcdeg in the frequency band of interest (3.5 to 7 mHz).

 \section{Data analysis}
The method we used to compute the coherence, and the phase differences is  described 
in Koopmans (1974). Briefly, let $A$ and $B$ be two time series of length $T$ and $\sin  A$, $\cos  A$, 
$\sin  B$, and $\cos  B$  be the sine and cosine amplitudes of the spectra for series $A$
and $B$. The power spectral densities, $P_A(\nu)$ and $P_B(\nu)$, the
co-spectral density, $C_{AB}(\nu)$, the quadrature spectral density,
$q_{AB}(\nu)$, and the complex cross-spectral density $P_{AB}(\nu)$
are defined as:

\begin{equation}
P_A(\nu)=\frac{T}{2}[\sin ^2A(\nu) + \cos ^2A(\nu)]
\end{equation}

\begin{equation}
P_B(\nu)=\frac{T}{2}[\sin ^2B(\nu) + \cos ^2B(\nu)]
\end{equation}

\begin{equation}
C_{AB}(\nu)=\frac{T}{2}[\sin A(\nu)\sin B(\nu)+\cos A(\nu)\cos B(\nu)]
\end{equation}

\begin{equation}
q_{AB}(\nu)=\frac{T}{2}[\sin A(\nu)\cos B(\nu)-\sin B(\nu)\cos A(\nu)]
\end{equation}

\begin{equation}
P_{AB}(\nu)=C_{AB}(\nu)-iq_{AB}(\nu)
\end{equation}

The coherence (the analogue of the linear correlation coefficient
between the two time series $A$ and $B$ in linear regression analysis) and the phase 
difference, $\Delta$$\phi_{AB}(\nu)$  between series $A$ and $B$ are given by:

\begin{equation}
Coh^2_{AB}(\nu)=\frac{\langle C_{AB}(\nu)\rangle ^2+\langle  q_{AB}(\nu)\rangle  ^2}{\langle  
P_A(\nu)\rangle  \langle P_B(\nu)\rangle  } =
\frac{|\langle P_{AB}(\nu)\rangle |^2}{ \langle  P_A(\nu)\rangle  \langle  P_B(\nu)\rangle  }
\end{equation}

\begin{equation}
\Delta\phi_{AB}(\nu)=\tan^{-1}(\frac{\langle  q_{AB}(\nu)\rangle  }{\langle  C_{AB}(\nu)\rangle  })
\end{equation}

where  $\langle $..$\rangle $ stands for a 5-bin smoothing of the respective functions. The length of the smoothing
is not too critical, but needs to be chosen adequately. If it is too short, the parameters 
are noisy from bin to bin, and if
it is too large the frequency resolution is lost. 
After several tests varying the length of the smoothing we concluded that 
5 bins is the best trade-off for this frequency range.

The errors for the phase difference are given by:

\begin{equation}
\epsilon_{\Delta\phi_{AB}}(\nu)=\sin ^{-1}(\sqrt{\frac{1-Coh_{AB}^2(\nu)}{(2n-2)Coh_{AB}^2(\nu)}}t_{2n-2}(\frac{\alpha}{2})),
\end{equation}

where {\em n} represents the equivalent degrees of freedom (EDF),
$t_{2n-2}(\alpha/2)$ the Student $t$-distribution,
and $\alpha$ the confidence level at which the errors are computed ($\alpha$=0.8 for phase differences).

 After computation of the power spectra and bivariate parameters for each pair of  
4-day time series, these are averaged to obtain four final power spectra (one for each of the 
three VIRGO/SPM channels and one for GOLF) and three sets of bivariate parameters (coherence 
and phase shift) between each of the VIRGO/SPM channels with GOLF. The power spectra and bivariate parameters
of 200 consecutive 4-day time series are thus averaged obtaining the results for the first  800-day time series
and this is repeated for the 85 subseries of 800 days.

The VIRGO signals show a peak (highest in the green channel) at just 5555 $\mu$Hz  (see Figure 1)
corresponding to 3 minutes. This is precisely the period of the calibration reference used by 
the Data Acquisition System (DAS) of VIRGO. This is an electronic artifact that in principle 
could contaminate the determination of the acoustic cut-off frequency but, as will be seen 
in the following sections, it is not a problem because the coherence and phase-shift are not 
affected by this artifact. Indeed, this signal is not present in the GOLF data, therefore it yields  a 
low value of the coherence 
at this frequency. The use of power spectra to determine the acoustic cut-off frequency requires special care in the 
treatment of this instrumental signal, but does not affect the coherence and phase-shift on which this work is based.

 Figure 2 shows two power spectrum densities (PSD), in the range 3000 to 7000 $\mu$Hz, computed 
from an example of two 800-day subseries (one in intensity and one in velocity).  We also show the 
resultant bivariate parameters.The coherence function has its maxima just where the p-mode maxima lie. This means that the coherence 
has the relative maxima at the frequencies in which both the intensity and velocity signals are 
coherent. In the pseudomode range  $\nu$ $ >$5000 $\mu$Hz the coherence function also shows some maxima 
because of the presence of pseudomodes in both signals. The visibility of the peaks in the coherence function 
is much higher than in the power spectra.

The bottom plot of Figure 2 shows the phase difference between the intensity and  the velocity signals. 
The exact value of the phase difference is at the frequencies where the coherence function has 
its maxima, usually close to the maxima of the phase difference function. The phase differences 
between I-V p-modes are close to the adiabatic value of $-$90 degrees (see Jim\'enez 2002) but beyond 
the acoustic cut-off frequency $\sim $5000 $\mu$Hz a change takes place: the phase differences 
decrease approximately linearly in this region. The frequency at which this different behavior of the 
phase difference takes place will be studied in Section~6.

\placefigure{figure1}

\placefigure{figure2}

\section{Determination of the acoustic cut-off frequency}
As explained in the introduction, a change in the value of the frequency spacing between p-modes 
and pseudomodes  is expected because of the increase in the large frequency separation around the 
acoustic cut-off frequency. 
To find where this frequency spacing change starts to take place,  a convenient definition of the 
acoustic cut-off frequency is derived and adopted in this section.

An exponentially modulated sine wave is fitted to the coherence function between 3500 and 
5500 $\mu$Hz to take into account the decreasing amplitude of p-modes in this frequency range. In Figure 3a 
the coherence function (black line) is plotted between 3500 and 6500 $\mu$Hz,  together with 
the modulated sine wave  (blue line) but extended to 6500 $\mu$Hz.

The shape of the solar signal in the pseudomode region (between 5000 $\mu$Hz and 6500 $\mu$Hz) is 
like a sine wave. Thus a single sine wave is fitted to the coherence function 
between 5000 and 6500 $\mu$Hz. This second fit is also plotted in Figure 3a (red line) and 
extended to  lower frequencies, down to 3500 $\mu$Hz. The interval from 5000 $\mu$Hz to 5500 $\mu$Hz 
is used in both fits because it is the interval where the acoustic cut-off frequency is expected to be 
found and also because, obviously, it is not possible to separate p-modes from pseudomodes before finding the 
acoustic cut-off frequency (in several tests this interval has been slightly changed and the same results were obtained).

The maxima of the three curves---coherence, modulated sine wave, and single sine wave---have been computed 
and plotted in Figure 3a, where black, blue, and red circles correspond to the frequencies of 
the respective maxima of these variables.  Up to 5000 $\mu$Hz the coherence function and the 
fitted modulated sine wave are in phase and their maxima have the same frequencies. No blue 
circles are visible below 5000 $\mu$Hz because they are over-plotted by the black ones.  Around 5000 $\mu$Hz the 
coherence function starts to shift to higher frequencies and the maxima of the fitted modulated sine wave are delayed 
with respect to those of the coherence function. Blue circles then become visible and the coherence shift 
increases with frequency, being out of phase (maxima of the coherence coincides with minima of the fitted modulated sine wave) between 
6000 and 6500 $\mu$Hz.

In the pseudomode region, just the opposite effect takes place. The coherence function and the 
fitted single sine wave go out of phase from higher to lower frequencies. From 6500 to around 5000 $\mu$Hz 
the coherence and the fitted single sine wave  are in phase and their maxima have the same frequencies. From 
around 5000  $\mu$Hz to lower frequencies the coherence function begins to delay with respect to the fitted single 
sine wave, and in fact they are out of phase (the maxima of the coherence coincides with the minima of the 
fitted function) between 4000 and 3500 $\mu$Hz (red circles are well visible).

To look in detail at these two opposing effects, the frequency differences between the maxima of Figure 3a are 
computed; that is, the frequency differences between the maxima of the coherence function and the maxima of 
the fitted modulated sine wave and between the maxima of the coherence function and the maxima of the single 
sine wave. In Figure 3b blue circles correspond to the former (p-modes) and red squares to the pseudomodes.

Blue circles have an approximately constant value close to zero in the frequency range where the p-mode coherence 
signal is in phase with the fitted function; that is, from 3500 up to $\sim$5000 $\mu$Hz. At this 
frequency the coherence shifts to higher frequencies and the values of the maxima differences increase up to 
38 $\mu$Hz at the frequencies where coherence and fitted function are out of phase.

The frequency differences between the maxima of the coherence and the maxima of the single sine wave fitted 
to the pseudomode range (red squares) are out of phase in the p-mode frequency range. From a value  of around 
40 $\mu$Hz at 3500 $\mu$Hz, these differences decrease with frequency up to, again, around 5000 $\mu$Hz. 
From this frequency the differences have an approximately constant value in the frequency range where the coherence
and the fitted function are in phase. This constant value is close to zero but with some dispersion,  probably 
because of the irregular
shape of the pseudomodes.

At this point, we define the acoustic cut-off frequency value as the crossing point 
of both frequency differences shown in Figure 3b. This crossing point is the frequency at which, hypothetically, the 
maxima of both fitted functions and the maxima of the coherence  coincide.

Looking around 5000 $\mu$Hz in Figure 3b, the blue points are always below the red ones for 
lower frequencies and always above for higher frequencies. The frequency interval between 
the transition of blue points from below to above the red points (indicated by a hexagon) 
is the frequency interval of the acoustic cut-off value. These two limits are the lower 
and upper limits of $\nu_{\rm ac}$. A more accurate determination of $\nu_{\rm ac}$ is
performed by fitting two parabolas (one for the red points and one for the blue points) in the interval  
from 4500 to 5500 $\mu$Hz and computing the their crossing frequency point. This is considered 
as the best feasible determination of $\nu_{\rm ac}$. 

  The acoustic cut-off frequency for low-degree modes found with this method is lower than indicated 
in previous observations in which a wide range of values were found between 5300  and 
5700 $\mu$Hz. All these previous determinations (Claverie, et al. 1981; Pall\'e et al. 1986, 1992; 
Duval et al. 1991; Fossat et al. 1992) were performed using ground-based observations, with different 
techniques, and with a poor visibility of the pseudomodes. A suggested explanation for these 
discrepancies could be the higher quality of the SOHO space-based data and the 
possibility that a lack of good visibility of the pseudomodes might require different techniques to consider the trace 
of the first pseudomodes as a p-mode signal.

  From the analysis of this section we can compute the evolution with time of the mean large frequency separation 
for p-modes and pseudomodes. From the exponentially modulated  sine wave fitted to the coherence function 
(p-mode range),  we can extract the separation for p-modes and from the sine wave fitted to the 
coherence function (the pseudomode range) we can extract the separation for pseudomodes. In Figure 4, we plot these 
differences (filled circles for p-modes and open squares for pseudomodes). An almost constant difference in the mean $\Delta\nu_{n,\ell}$
is observed for p-modes and pseudomodes, with a difference around 2 $\mu$Hz. The higher dispersion of $\Delta\nu_{n,\ell}$
for pseudomodes before 2002--2003 is probably the influence of the rising part of the solar activity cycle which makes 
the pseudomode range  noisier  where the S/N ratio is much lower than that for the p-modes.

\rm

\placefigure{figure3}

\placefigure{figure4}

\section{The acoustic cut-off frequency and the solar activity cycle}
The final values of the acoustic cut-off frequency, computed from the 800-day time series 
described in Section 3,  are shown in Figure 5 for the three VIRGO/SPM photometers and for GOLF.
The values of the acoustic cut-off obtained as the crossing point of the two parabolic segments 
(Figure 3b) are represented by black dots whereas the two limits (the frequency interval
between the transition of blue points from below to above the red points in Figure 3b) are shaded in gray.  

The acoustic cut-off for the three channels of VIRGO/SPM (red, green, and blue), together with the 
solar radio flux integrated as the time series used for the $\nu_{\rm ac}$ computations, are plotted 
in Figure 6, where a clear correlation between $\nu_{\rm ac}$ and the solar activity cycle can be 
seen. We computed different correlation indices between the acoustic cut-off frequency of the red, green, and blue 
VIRGO/SPM channels, and the solar radio flux:  the Pearson coefficient correlation 
(P), the Spearman rank correlation (S) and its two-sided significance P$_s$ (probability of having a 
null correlation). These correlations are shown in Table 2, from which we may conclude that there 
is a high positive correlation between the $\nu_{\rm ac}$ and the magnetic activity. 

The p-mode frequency shift produced by magnetic activity cannot explain the variation observed in 
$\nu_{\rm ac}$, as the amount of this shift is of the order of 0.4 $\mu$Hz (Broomhall et al. 2009; 
Salabert et al. 2011), much smaller than our present bin width (2.89 $\mu$Hz), and the obtained 
amplitude of the  $\nu_{\rm ac}$ variation is around 100 $\mu$Hz.

\placefigure{figure5}
\placefigure{figure6}

\begin{table}
\caption{\label{tab:correlations}Results from the correlations between the acoustic cut-off frequency 
and the solar radio flux, The Pearson coefficient correlation (P), the Spearman rank correlation (S) 
and its two-sided significance P$_s$(probability of having a null correlation) for the red, green, and blue VIRGO/SPM 
channels.}
\begin{center}
\begin{tabular}{llll}
\hline
Channel & P  & S &P$_s$\\
\hline
Red    & 0.86 & 0.84 & 1.67 E$^{-23}$ \\
Green & 0.85 & 0.88 & 1.15 E$^{-28}$\\
Blue    & 0.77 & 0.75 & 1.79 E$^{-16}$\\
\hline
\end{tabular}
\end{center}
\end{table}

As mentioned in the introduction we also investigate how the I-V phase differences
could be affected by the transition between p-modes and pseudomodes. Figure 7 shows the phase differences between 
4000 $\mu$Hz and 6000 $\mu$Hz for three different time series. As explained in
Section 4, the I-V phase differences between p-modes does not correspond to the adiabatic 
value ($-$90 degrees) but is close to it, reaching the higher values close to the acoustic 
cut-off frequency.  In the pseudomode regions the phase differences decrease almost linearly, 
giving a parabolic shape to the I-V phase differences in this frequency interval 
(4000 $\mu$Hz to 6000 $\mu$Hz). The black points in each of the three plots of Figure 7 are 
the values of the phase difference function  at the frequencies where the coherence function 
has its maxima; i.e., the phase differences between p-modes and between pseudomodes. 
The two parabolic lines in each plot are two fitted parabolas, one using the phase difference 
values (black dots) and the other using the whole phase difference function. The purpose of 
these fits is to know the position of the maxima of these two parabolas.
From a first inspection, it is clear that the maxima of the first plot (a) and third 
plot (c)  are similar, whereas the maxima of the middle plot (b) seem to be at higher frequencies. 
The first and third plots correspond to time series close to the minimum of solar activity, a) in 
1996 and b) in 2009 and the middle plot correspond to a time serie close to  the maximum
of solar activity (2001). This analysis have been done for the 85 time series and the maxima of 
the two parabolas computed is plotted in Figure 8. The different colors correspond to the 
red, green, and blue channels of VIRGO/SPM and the black line is the solar radio flux integrated 
in the same subseries. Although the values plotted in Figure 8 cannot be considered as a direct 
measurement of the acoustic cut-off frequency, the phase difference function at these frequencies is related 
to the presence of pseudomodes, and therefore with $\nu_{\rm ac}$; hence, it follows the same 
correlation with the solar activity cycle as the values of $\nu_{\rm ac}$ computed previously (Figure 6).

\placefigure{figure7}

A detailed analysis of Figure 8 suggests a delay between the solar cycle and the position 
of the maxima of the parabolic fits to the phase difference functions. By plotting these values
 as a function of the solar radio flux we obtain the results shown in Figure 9 (for the green channel). Black symbols 
 correspond to the maxima of Figure 8 for the fit to the whole phase difference function. The filled
 circles correspond to the ascending part of the cycle, and the open squares to the descending one. The 
 maxima of the phase difference do not have a linear dependence on the solar cycle and the path 
 in the ascending and descending part of the cycle are different. This is a ``hysteresis'' cycle 
 which was already found in other p-mode parameters as the frequency shifts (e.g., Jimenez-Reyes et al.\  
 1998). On the other hand, we also plot the acoustic cut-off as a function of the solar radio flux (in red), 
which follows a linear relation with the solar cycle (the straight red line is the resultant fit). 
In Table 3 we show the slope of the straight lines for the three channels of VIRGO/SPM.

\placefigure{figure8}

\placefigure{figure9}

\begin{table}
\caption{\label{tab:correlations}Slope of the linear fits to $\nu_{\rm ac}$ as function of solar radio flux.}
\begin{center}
\begin{tabular}{ll}
\hline
Channel & Slope (W/$m^2$/Hz/$\mu$Hz) \\
\hline
Red    & 0.759 \\
Green & 0.680  \\
Blue    & 0.654 \\
\hline
\end{tabular}
\end{center}
\end{table}

\section{Conclusions}
In this work we have determined the acoustic cut-off frequency of the Sun, $\nu_{\rm ac}$, using pseudomode properties,
in particular, the increase in the large separation around $\nu_{\rm ac}$ that set the limit 
between p-modes and pseudomodes. Instead of power spectra we used the coherence function obtained 
with bivariate analysis between simultaneous intensity and velocity time series. Our results show 
that the value of $\nu_{\rm ac}$ is lower than the theoretical value and also lower than previous 
determinations in which power spectra were used without taking into account the influence of the 
pseudomodes. The acoustic cut-off frequency has been determined along the solar activity cycle
during the 15 years of SOHO measurements. The results show a clear correlation of $\nu_{\rm ac}$ 
with magnetic activity, the mean value being  around 5000 $\mu$Hz. The variation between maximum and 
minimum is around 100--150 $\mu$Hz. This relation between  $ \nu_{\rm ac}$ and magnetic activity is linear.

The exact determination of the solar $\nu_{\rm ac}$ and its variation during the solar cycle 
is important not only intrinsically, but also 
for its asteroseismic implications. On the one hand, scaling relations from solar values are widely 
generalized in asteroseismology and these relations uses the maximum of the p-mode hump, which
is related with $\nu_{\rm ac}$. On the other hand, magnetic activity cycles have begun to be
detected in other stars by means of asteroseismology (Garc\'\i a et al. 2010). Thus, a better
knowledge of the behavior of the cut-off frequency would improve our understanding of other stars.

To apply the method used in this research to other stars, simultaneous intensity and velocity 
observations are required. The KEPLER satellite provides photometric measurements of thousands of
solar-like stars, but from different ground-based campaigns such as the forthcoming SONG (Stellar
Observations Network Group), simultaneous velocity data could be obtained at least for a 
couple of bright stars observed by KEPLER to try to determine the probable existence of
pseudomodes and find the acoustic cut-off frequency in solar-like stars.   

The dependence of  $\nu_{\rm ac}$ with magnetic activity is also verified by the 
position of I-V phase difference maxima in the interval from 4000 $\mu$Hz to 6000 $\mu$Hz, indicating 
a transition from p-modes to pseudomodes. This transition  is also correlated with the solar 
activity cycle in the same sense but with a hysteresis cycle pattern that could be a
manifestation of surface effects or due to time-delayed responses to a single phenomenon 
located deeper in the Sun.

\acknowledgments
This work was funded by the Spanish grant AYA2004-04462 and AYA2010-17803 of the Ministry of 
Science. RAG acknowledges the support of the CNES grant at the SAP CEA/Saclay. 
The GOLF and VIRGO instruments  benefit from the quiet and well run SOHO platform built by 
Matra--Marconi Space. SOHO is an international collaboration program of the European Space Agency
(ESA) and the National Aeronautics Space Administration (NASA). VIRGO and GOLF represent the cooperative 
efforts of many individual scientists and engineers at several institutes in Europe and the USA to 
whom we are deeply indebted.

\clearpage

\begin{figure}[!htb]
\centerline{%
\begin{tabular}{c@{\hspace{1pc}}c}
\includegraphics[width=30pc,angle =90]{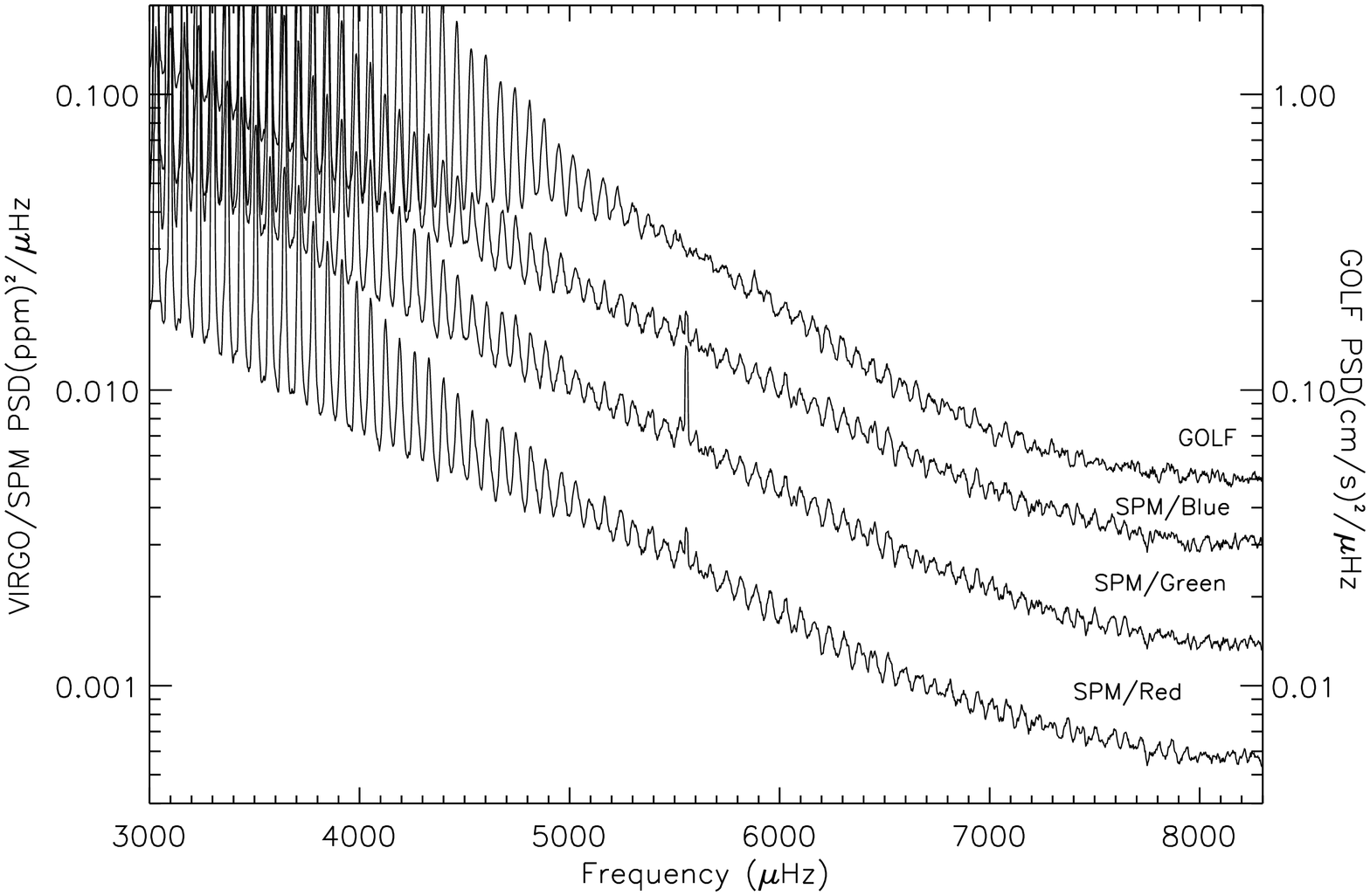}
\end{tabular}}
\caption{ Smoothed power spectra showing the clear periodic structure of pseudomodes for the signals used in this work. 
Bottom to top: The red, green, and blue channels of VIRGO/SPM, and the GOLF spectra. }
\end{figure} 

\clearpage

\begin{figure}[!ht]
\centerline{%
\begin{tabular}{c@{\hspace{1pc}}c}
\includegraphics[width=50pc,angle =90]{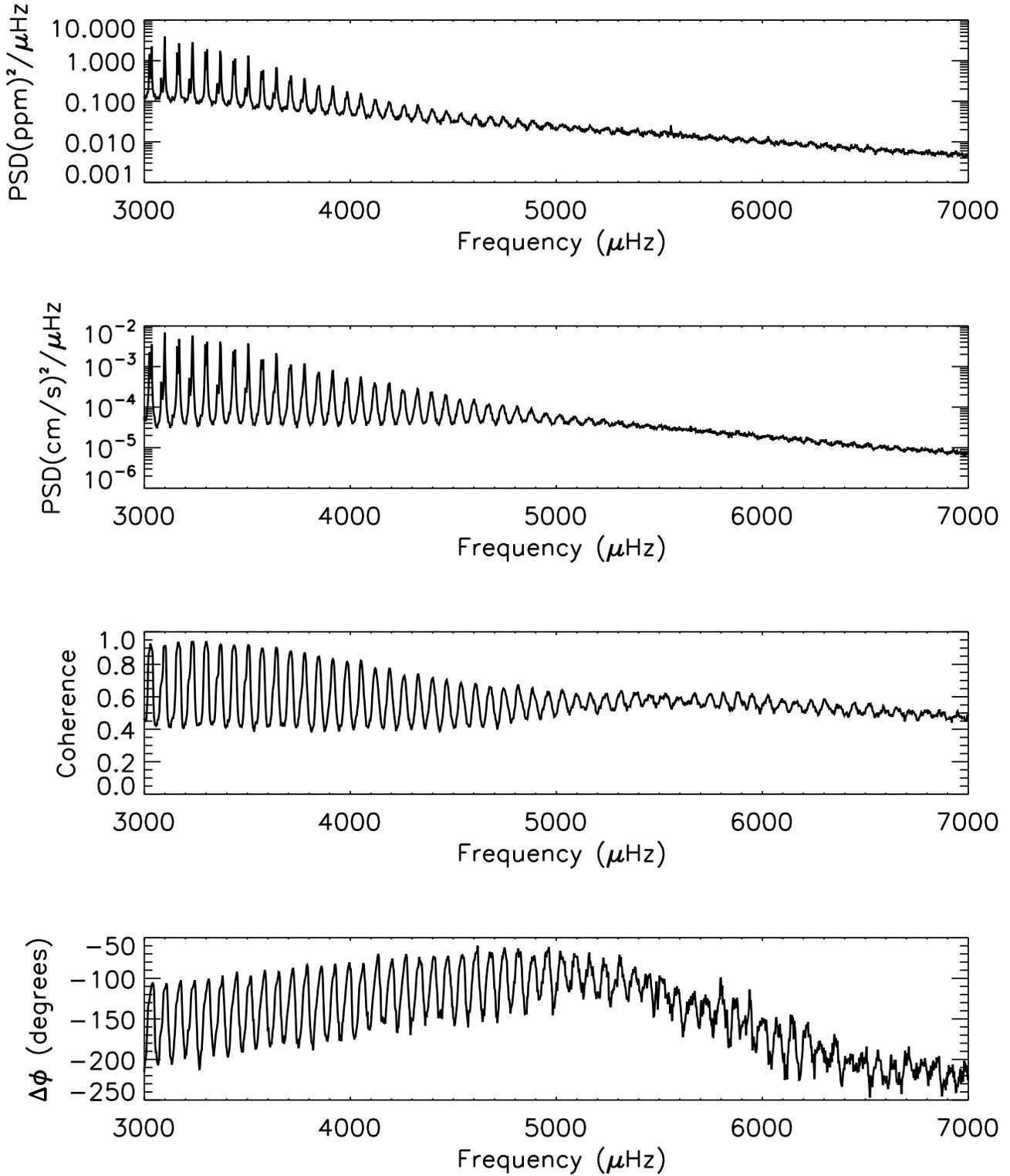}
\end{tabular}}
\caption{Intensity and velocity power spectra of one of the time series used and 
the corresponding bivariate parameters between both time series, coherence and phase difference.  }
\end{figure} 

\begin{figure}[!htb]
\centerline{%
\begin{tabular}{c@{\hspace{0pc}}c}
\includegraphics[width=30pc,angle=90]{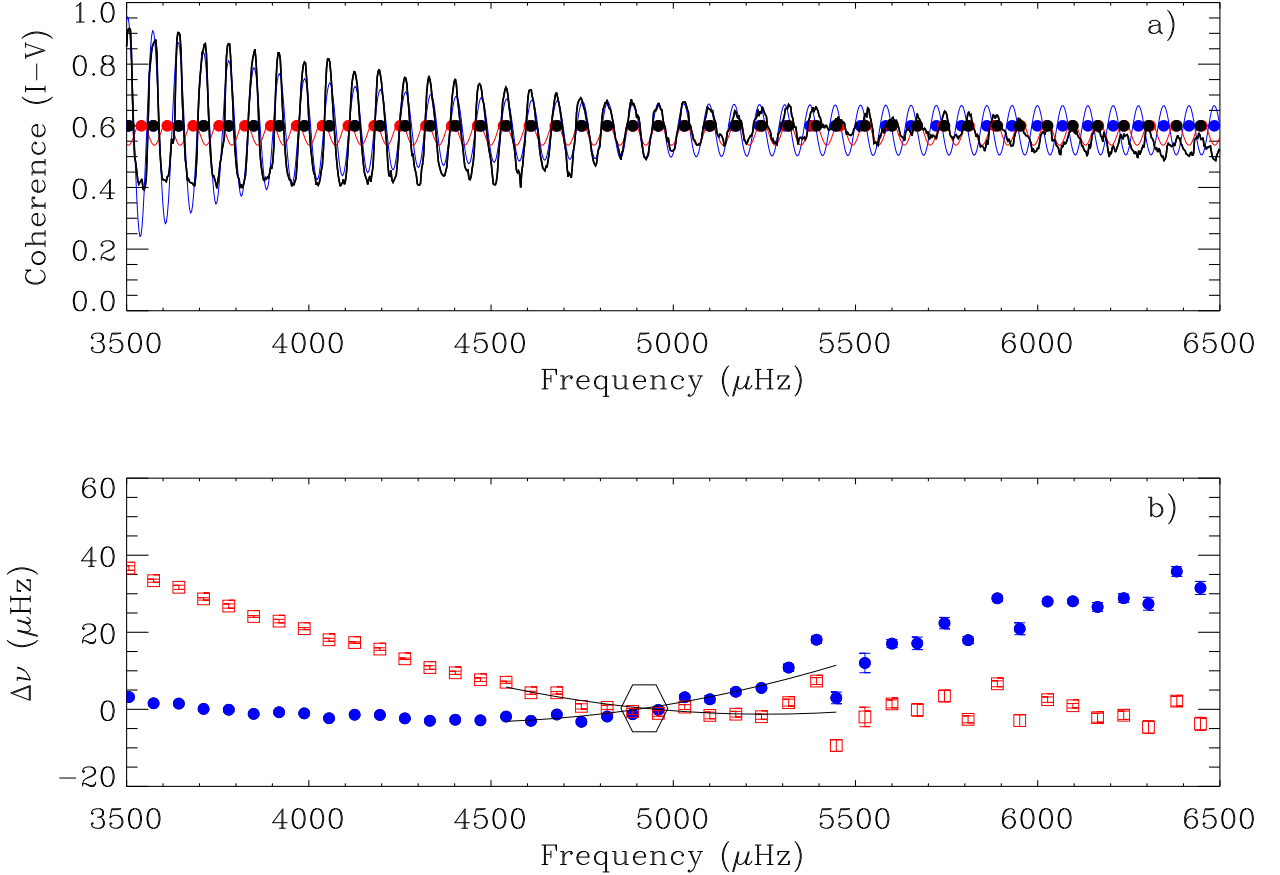}
\end{tabular}}
\caption{ a) Black line: coherence function. Blue line: Exponentially decaying sine wave fitted to 
the coherence function between 3500 $\mu$Hz and 5500 $\mu$Hz (end of the p-mode range) and 
extended  to 6500 $\mu$Hz.  Blue and black filled circles are the maxima of the fitted sine wave and 
coherence function respectively. Note how the coherence shifts to higher frequencies from 
$\sim$5000 $\mu$Hz onwards. Red line: Single sine fitted to the frequency range between 5000 $\mu$Hz and 
6500 $\mu$Hz (pseudomode range) and plotted extended to 3500 $\mu$Hz . Red and black filled circles 
are the maxima of the fitted sine and the coherence function.
b) Blue circles: frequency differences between the maxima of the coherence function and the fitted 
(3500--5500 $\mu$Hz) exponentially modulated sine wave. Red squares: frequency differences between 
the maxima of the coherence function and the fitted sine (5000--6500 $\mu$Hz).  Two parabola 
segments are fitted to the central part to get an estimation of $\nu_{\rm ac}$ }
\end{figure}

\clearpage

\begin{figure}[!ht]
\centerline{%
\begin{tabular}{c@{\hspace{1pc}}c}
\includegraphics[width=30pc,angle=90]{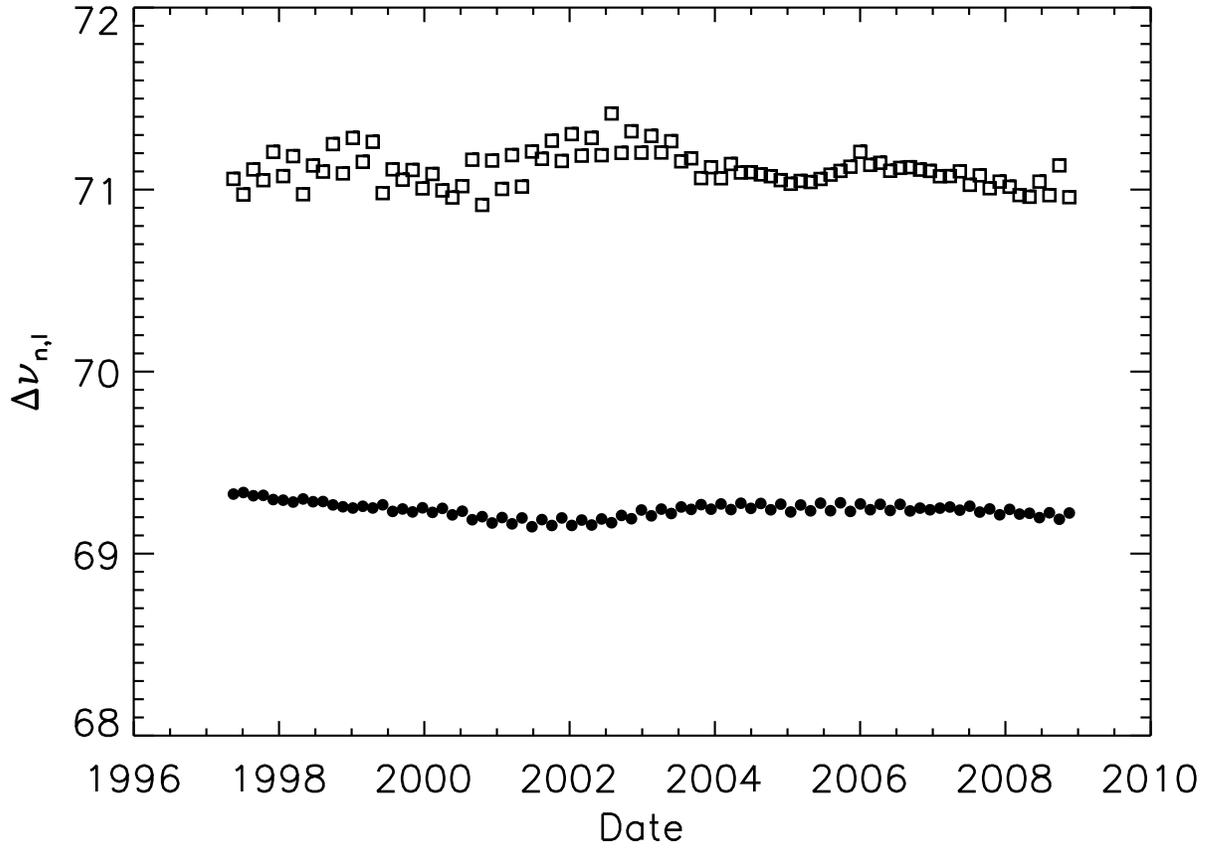}\end{tabular}}
\caption{ Time variation in the mean frequency separation  
$\Delta\nu_{n,\ell}$=$\nu_{\rm n,\ell}$-$\nu_{n-1,\ell}$  for p-modes (filled circles) 
and for pseudomodes (open squares) obtained from the exponentially modulated  sine wave 
fitted to the p-mode range and from the sinusoidal fit to the pseudomode range. }
\end{figure}

\begin{figure*}
\begin{center} 
\includegraphics[scale=0.4,angle=90]{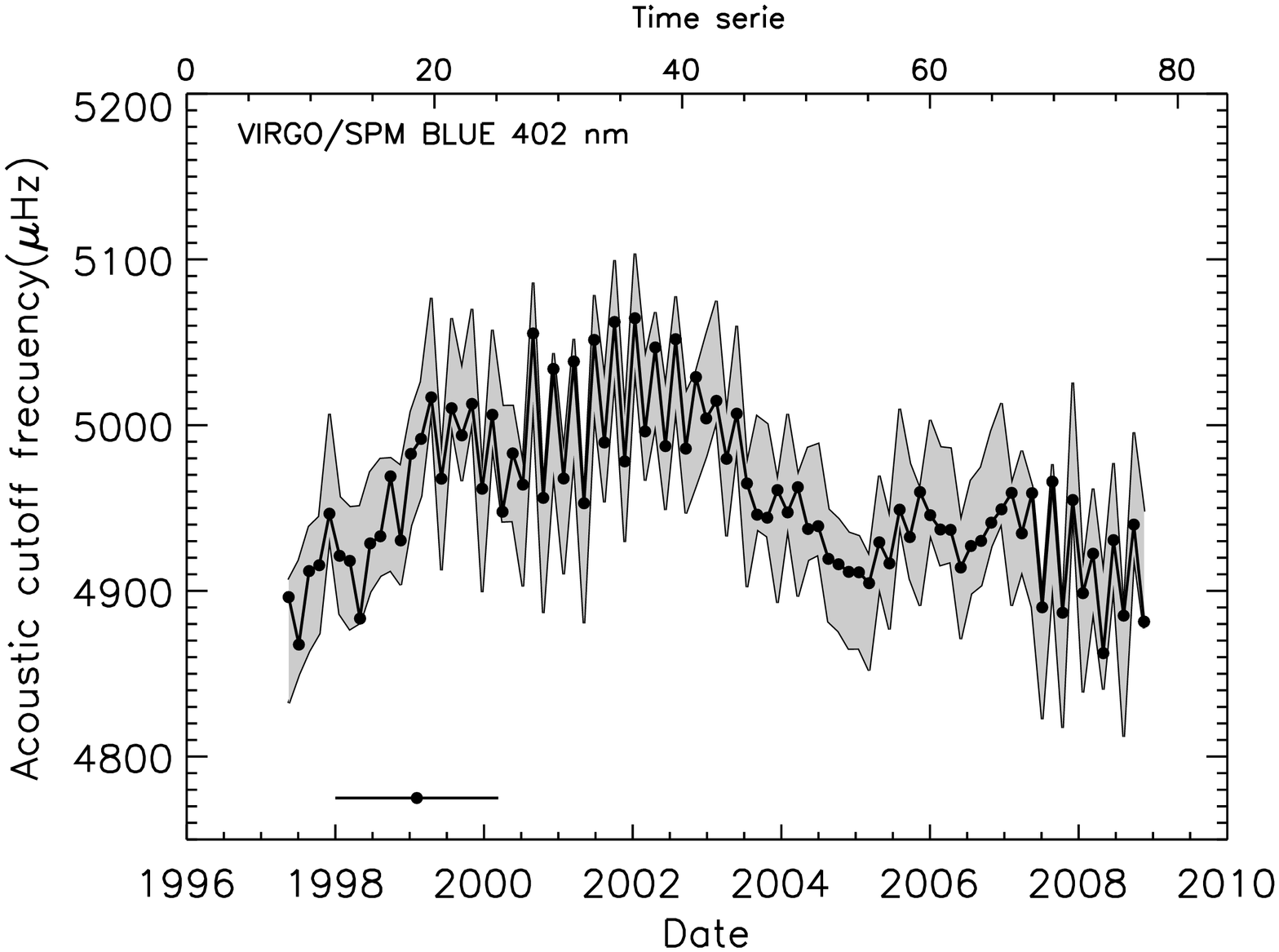} 
\includegraphics[scale=0.4,angle=90]{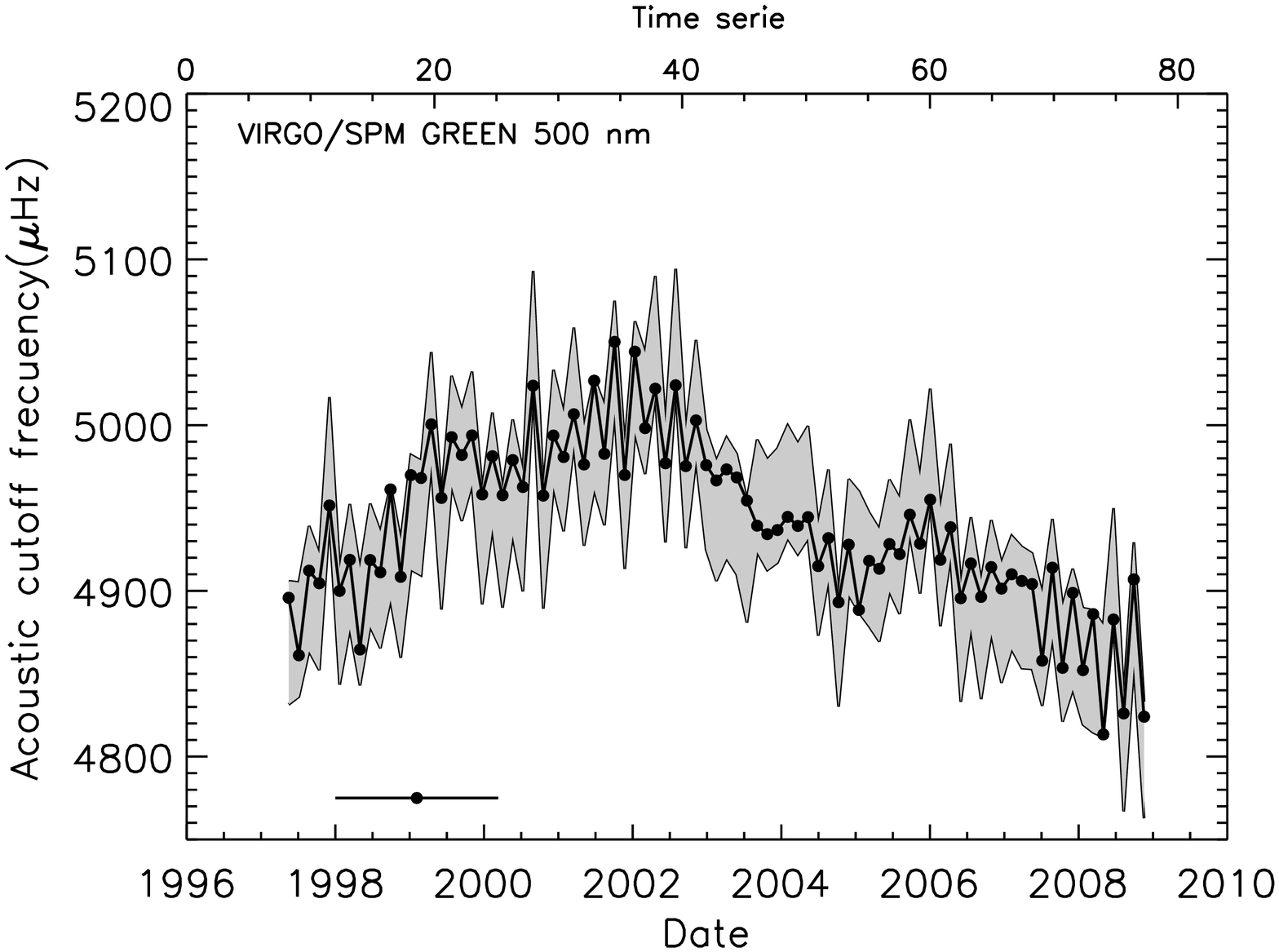} 
\includegraphics[scale=0.4,angle=90]{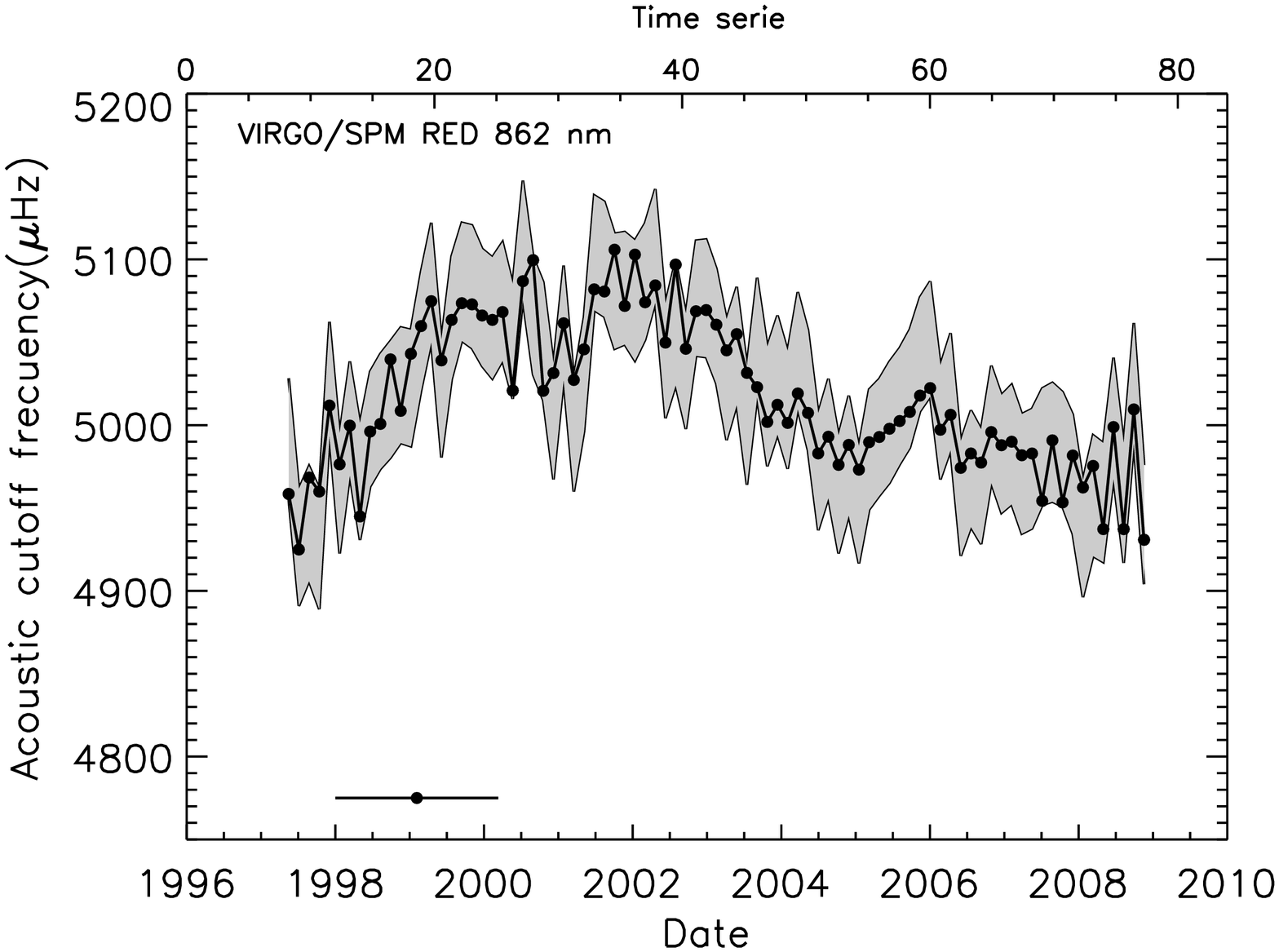} 
\end{center} 
\caption{The variation of the acoustic cut-off frequency for the blue, green, and red channels of 
VIRGO/SPM  for the 85 times series used in this research (the horizontal 
line centered at 1999 is the time span of the time series). Black points are the values of 
$\nu_{\rm ac}$ as obtained as the crossing point of the two parabolic segments as explained 
in the text and the gray areas are their two limits.  } 
\end{figure*} 
\clearpage

\begin{figure}[!ht]
\centerline{%
\begin{tabular}{c@{\hspace{1pc}}c}
\includegraphics[width=30pc,angle=90]{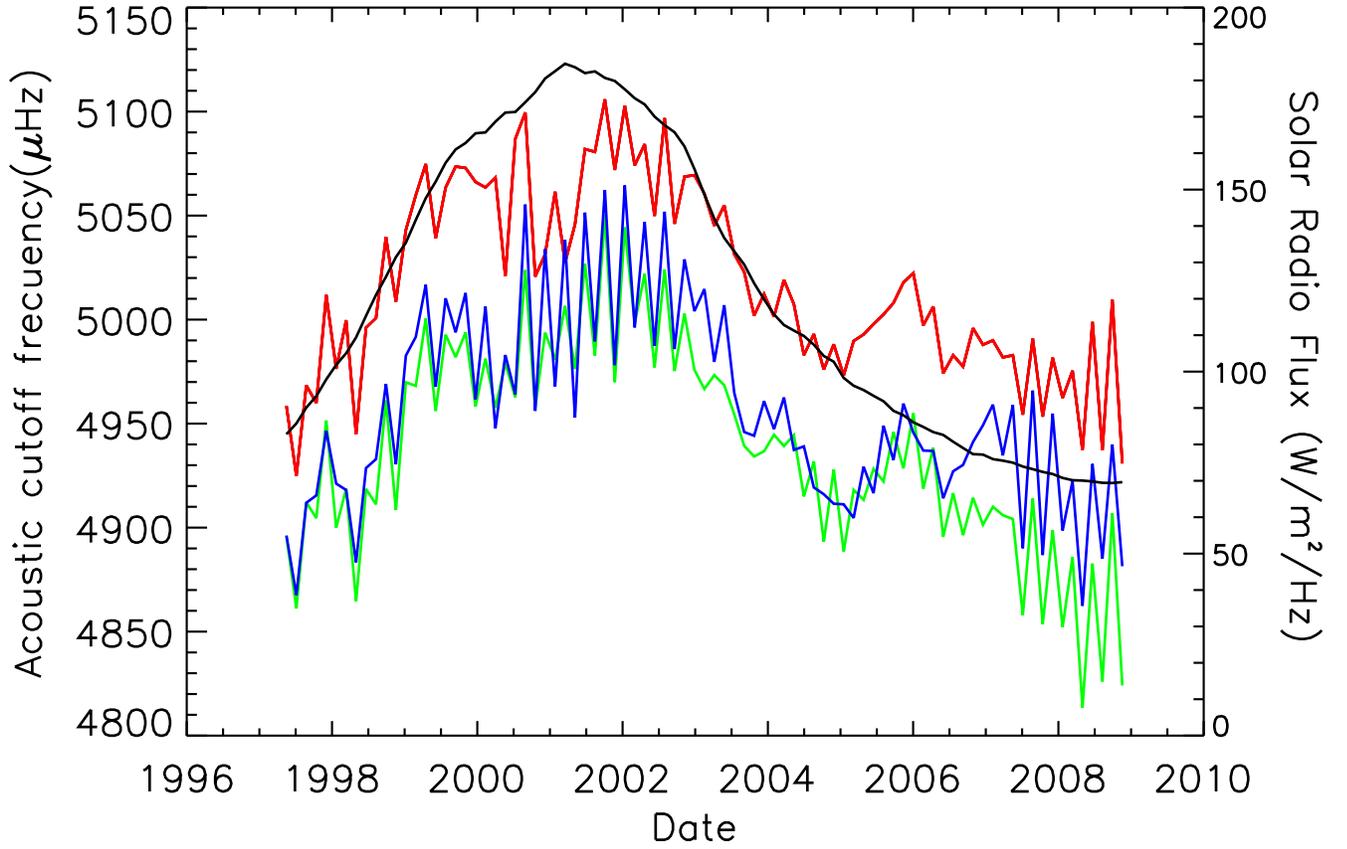}\end{tabular}}
\caption{ The acoustic cut-off for the red, green, and blue channels of VIRGO/SPM 
together with the radio flux integrated as the time series used for the $\nu_{\rm ac}$, average 
of 800 days shifting 50 days. $\nu_{\rm ac}$ is highly correlated with the solar activity cycle (see Table 2). }
\end{figure}

\clearpage

\begin{figure}[!t]
\centerline{%
\begin{tabular}{c@{\hspace{1pc}}c}
\includegraphics[width=30pc,angle=90]{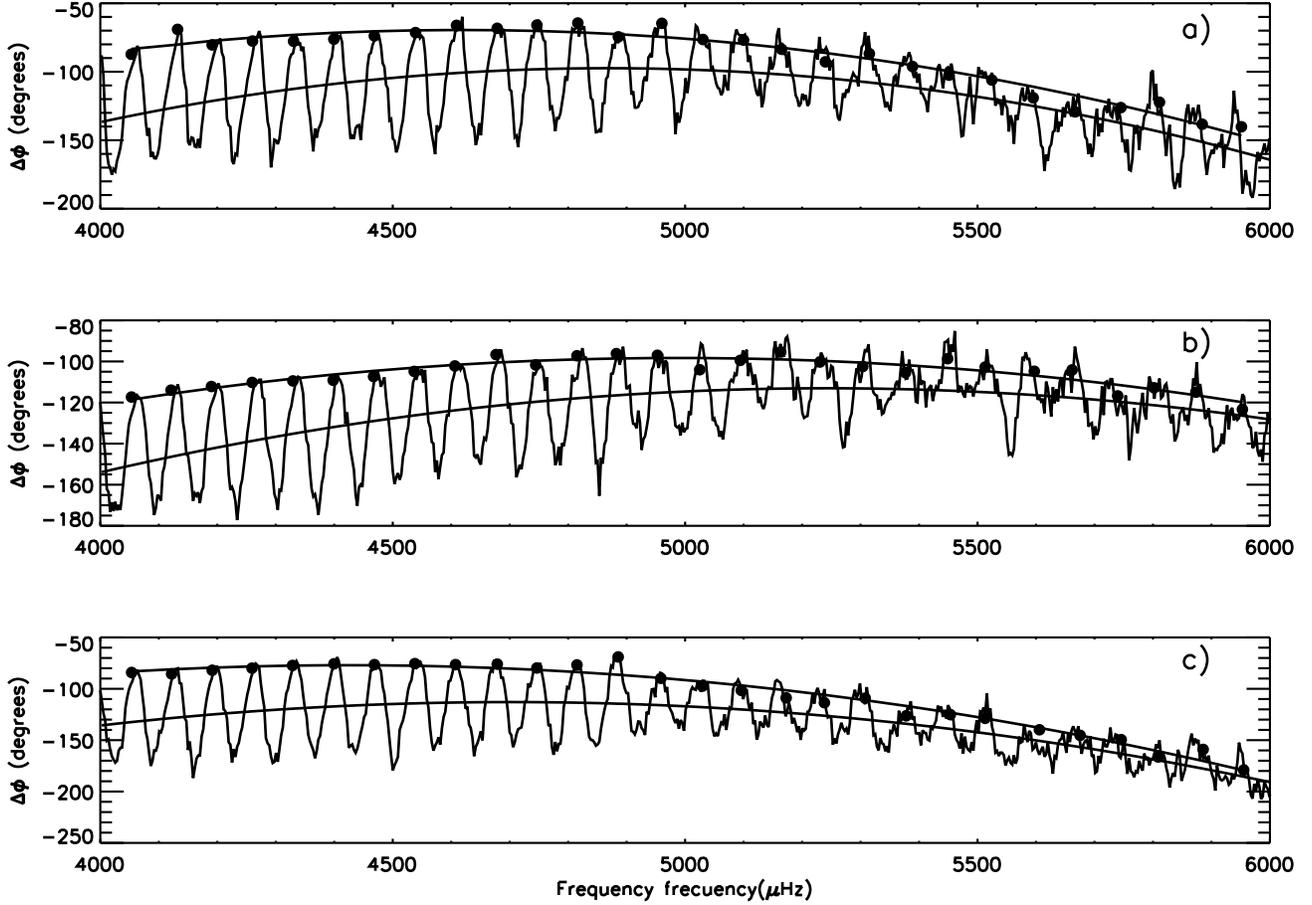}
\end{tabular}}
\caption{I-V phase differences in the region 4000 to 6000 $\mu$Hz for three time 
series taken at different epochs:  a) and c)  in 1996 and 2009 respectively (around the minimum of the
solar cycle) and b) in 2002 (around the maximum of the solar cycle). Black points are the values of 
phase differences just at the frequencies where the coherence function has its maxima. In 
each plot there are two fitted parabolas, one  to the black points and the other to the whole 
phase difference function (see text for more details). }
\end{figure}

\clearpage

\begin{figure}[!ht]
\centerline{%
\begin{tabular}{c@{\hspace{1pc}}c}
\includegraphics[width=30pc,angle=90]{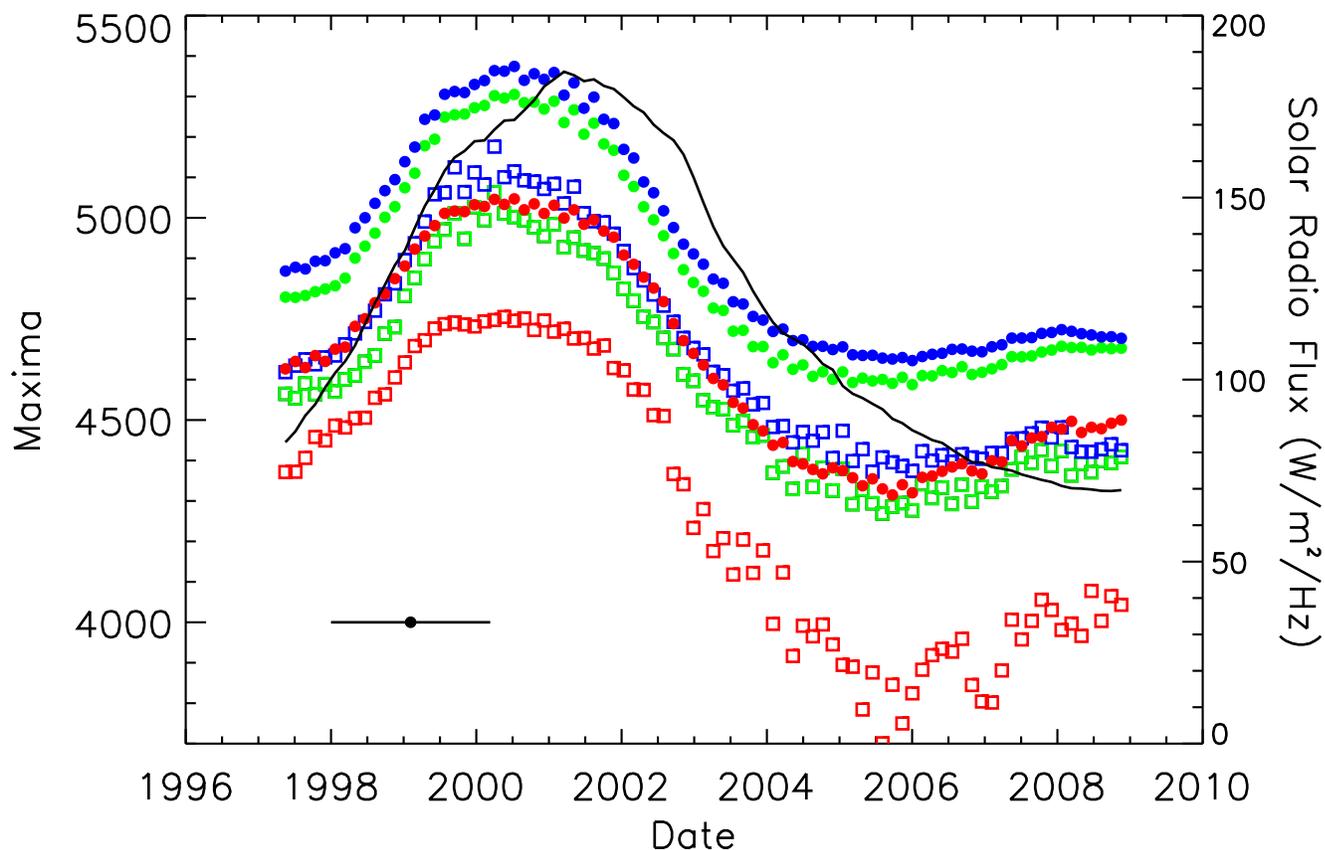}\end{tabular}}
\caption{Maxima of the two parabolic fits to the phase difference functions.
The different colors correspond to the three channels
of VIRGO/SPM, red, green, and blue. Solid points correspond 
to the maxima of the fitted parabola at the whole phase difference function and open squares to the 
exact values of the phase differences (the black points in Figure 7). The phase differences also verify 
that the $\nu_{\rm ac}$ is correlated with the solar activity cycle. }
\end{figure}

\begin{figure}[!ht]
\centerline{%
\begin{tabular}{c@{\hspace{1pc}}c}
\includegraphics[width=30pc,angle=90]{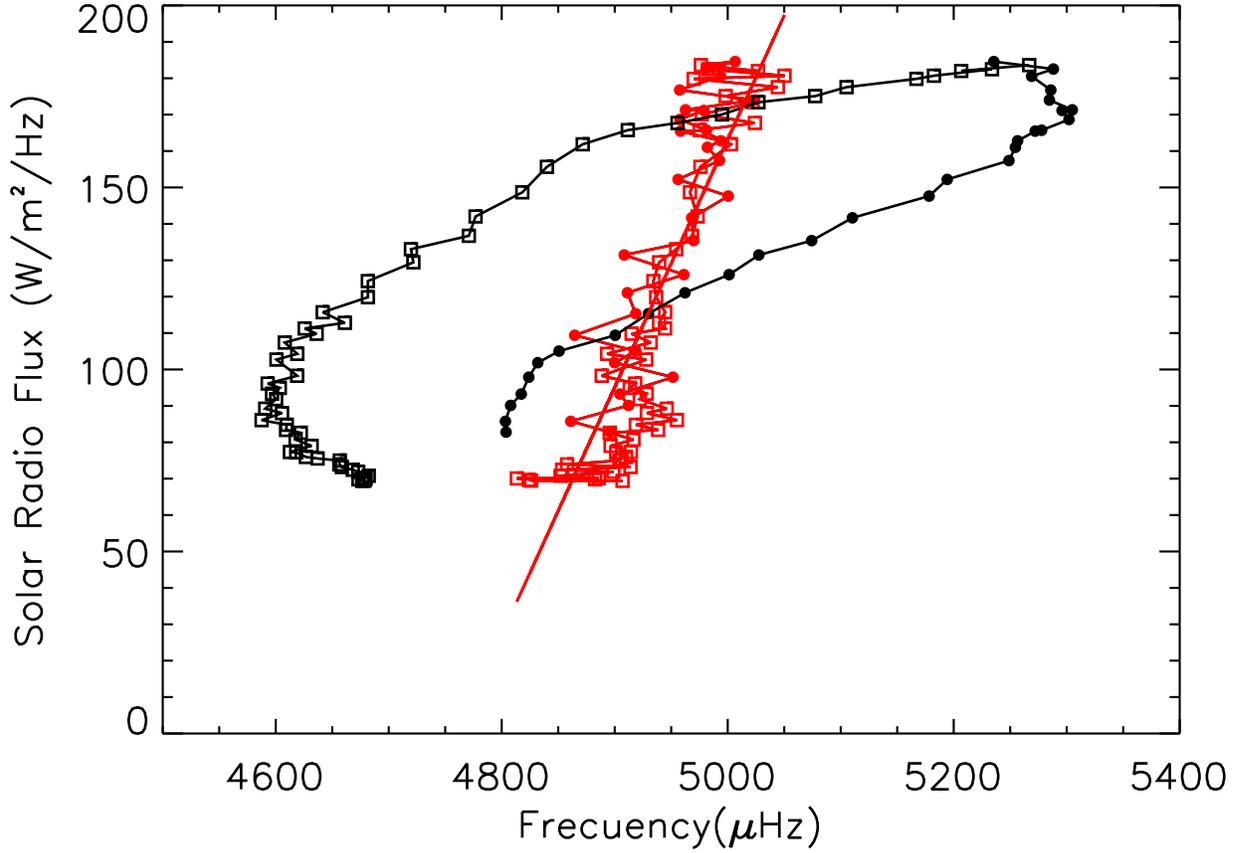}\end{tabular}}
\caption{Maxima of I-V phase differences and $\nu_{\rm ac}$ as a function of solar radio flux. Black 
symbols correspond to the maxima of Figure 8 for the fit to the whole phase difference function,
filled circles for the ascending part of the cycle and  open squares for the descending one. The maxima of 
the phase difference do not have a linear dependence on the solar cycle and the path in the 
ascending and descending parts of the cycle are different, that is, a ``hysteresis'' cycle. The acoustic 
cut-off as function of solar radio flux  is plotted in red (filled circles correspond to the ascending part 
of the cycle and open squares to the descending one), $\nu_{\rm ac}$ varies linearly with solar magnetic activity. }
\end{figure}

\end{document}